\documentstyle[12pt]{article}
\newcommand\Kahler{K\"{a}hler }

\catcode`\@=11
\def\marginnote#1{}
\def\ifmath#1{\relax\ifmmode #1\else $#1$\fi}

\def\bold#1{\setbox0=\hbox{$#1$}%
     \kern-.025em\copy0\kern-\wd0
     \kern.05em\copy0\kern-\wd0
     \kern-.025em\raise.0433em\box0 }

\def\GENITEM#1;#2{\par\vskip6pt \hangafter=0 \hangindent=#1
   \Textindent{$ #2$ }\ignorespaces}

\newcount\hour
\newcount\minute
\newtoks\amorpm
\hour=\time\divide\hour by60
\minute=\time{\multiply\hour by60 \global\advance\minute by-
\hour}
\edef\standardtime{{\ifnum\hour<12 \global\amorpm={am}%
    \else\global\amorpm={pm}\advance\hour by-12 \fi
    \ifnum\hour=0 \hour=12 \fi
    \number\hour:\ifnum\minute<100\fi\number\minute\the\amorpm}}
\edef\militarytime{\number\hour:\ifnum\minute<100\fi\number\minute}
\def\draftlabel#1{{\@bsphack\if@filesw {\let\thepage\relax
  \xdef\@gtempa{\write\@auxout{\string
    \newlabel{#1}{{\@currentlabel}{\thepage}}}}}\@gtempa
    \if@nobreak \ifvmode\nobreak\fi\fi\fi\@esphack}
     \gdef\@eqnlabel{#1}}
\def\@eqnlabel{}
\def\@vacuum{}
\def\draftmarginnote#1{\marginpar{\raggedright\scriptsize\tt#1}}
\def\draft{\oddsidemargin -.5truein
        \def\@oddfoot{\sl preliminary draft \hfil
        \rm\thepage\hfil\sl\today\quad\militarytime}
        \let\@evenfoot\@oddfoot \overfullrule 3pt
        \let\label=\draftlabel
        \let\marginnote=\draftmarginnote

\def\@eqnnum{(\theequation)\rlap{\kern\marginparsep\tt\@eqnlabel}%
\global\let\@eqnlabel\@vacuum}  }
\def\preprint{\twocolumn\sloppy\flushbottom\parindent 1em
        \leftmargini 2em\leftmarginv .5em\leftmarginvi .5em
        \oddsidemargin -.5in    \evensidemargin -.5in
        \columnsep 15mm \footheight 0pt
        \textwidth 250mmin      \topmargin  -.4in
        \headheight 12pt \topskip .4in
        \textheight 175mm
        \footskip 0pt

\def\@oddhead{\thepage\hfil\addtocounter{page}{1}\thepage}
        \let\@evenhead\@oddhead \def\@oddfoot{} \def\@evenfoot{}
}
\def\titlepage{\@restonecolfalse\if@twocolumn\@restonecoltrue\o
necolumn
     \else \newpage \fi \thispagestyle{empty}\c@page\z@
        \def\thefootnote{\fnsymbol{footnote}} }
\def\endtitlepage{\if@restonecol\twocolumn \else  \fi
        \def\thefootnote{\arabic{footnote}}
        \setcounter{footnote}{0}}  %\c@footnote\z@ }
\catcode`@=12
\relax
\def\be{\begin{equation}}
\def\ee{\end{equation}}
\def\bea{\begin{eqnarray}}
\def\eea{\end{eqnarray}}

\def\mst11{m_{\;\widetilde{t}_{1}}}

\def\mst22{m_{\;\widetilde{t}_{2}}}
\def\mst12{m_{\;\widetilde{t}_{1,2}}}

\def\msb11{m_{\;\widetilde{b}_{1}}}
\def\msb22{m_{\;\widetilde{b}_{2}}}
\def\msb12{m_{\;\widetilde{b}_{1,2}}}

\def\mwidetilde2{\widetilde{m}^{2}}

\relax
\begin{document}
\input epsf

\topmargin-2.5cm
%\draft
%\preprint
%
\begin{titlepage}
\begin{flushright}
CERN-TH/2002-307\\
hep--ph/0210427 \\
\end{flushright}
\vskip 0.3in
\begin{center}
{\Large\bf Quintessence and the Underlying}
\\
\vskip 0.2cm
{\Large\bf  Particle Physics Theory}

\vskip .5in
{\large\bf D.J.H. Chung$^{(1)}$\footnote{\baselineskip=16pt 
E-mail: {\tt daniel.chung@cern.ch}},}
{\large\bf L.L. Everett$^{(1)}$\footnote{\baselineskip=16pt 
E-mail: {\tt Lisa.Everett@cern.ch}},}
 {\bf and} 
{\large \bf A. Riotto$^{(2)}$\footnote{\baselineskip=16pt Email:
{\tt antonio.riotto@pd.infn.it}}}

\vskip 1cm
$^{(1)}$ {\it CERN Theory Division,
CH-1211 Geneva 23, Switzerland}
\vskip 0.5cm
$^{(2)}$
{\it INFN, Sezione di Padova, via Marzolo 8, I-35131, Padova, Italy}

\end{center}
\vskip2cm
\begin{center}
{\bf Abstract}
\end{center}
\begin{quote}
At present we know nothing about the nature of the dark energy 
accounting for about $70\%$ of the energy density of the Universe.
One possibility is that the dark energy is provided by an extremely
light field, the quintessence, rolling down its potential.
%In this paper we make two simple observations. 
Even though 
the underlying particle theory responsible
for the present quintessential behaviour of our Universe is 
unknown, such a 
theory is likely to have contact  with supersymmetry, supergravity or  
(super)string theory. In these
theories, there are plenty of scalar fields (moduli) 
which are gravitationally coupled to
all the other degrees of freedom and have vacuum expectation values of
the order of the Planck scale. 
%They generically go under the name 
%of moduli.
We point out that,  in theories
which allow a consistent embedding of quintessence, 
the generic gravitational interaction of the
moduli fields with the quintessence field gives rise to a contribution
to the energy density from the moduli fields 
of the order of the critical energy density of
the universe today.  Furthermore, the interaction contribution can
generically enhance the negativity of the equation of state.
\end{quote}
\vskip1.cm
\begin{flushleft}
%CERN-TH/96-242\\
October 2002 \\
\end{flushleft}

\end{titlepage}
\setcounter{footnote}{0}
\setcounter{page}{0}
\newpage
%
% BODY
\baselineskip=18pt
\noindent

{\bf 1.}~~There is increasing evidence that the energy density of
(baryonic plus dark) matter in the universe is smaller than the
critical density \cite{reviewparameters}.  If the universe is flat, as
predicted by the most natural inflation models \cite{reviewinf} and
confirmed by the recent measurements of the cosmic microwave
background anisotropies \cite{reviewcmb}, an additional dark energy
density component is necessary to account for $\Omega_0=1$.  The data
indicates the dark energy component possesses negative pressure and
makes up about 70 \% of the energy in the present universe.
%An obvious candidate is
%represented by the cosmological constant, whose equation of state is
%$\rho=-p$ (where $\rho$ is its energy density and $p$ is its
%pressure). If this is the option chosen by Nature, there is a
%Herculean task to explain why the energy of the vacuum is of the order
%of $(3\times 10^{-3}\:{\rm eV})^4$ much smaller than what is expected
%from our present understanding of fundamental fields of nature.
%Another possibility invokes a mixture of cold dark matter and
%quintessence \cite{quint}, a slowly-varying, spatially inhomogeneous
%component with equation of state $p_Q=w_Q\rho_Q$, with $-1< w_Q\leq
%0$.  The role of quintessence may be played by any scalar field $Q$
%which is slowly rolling down its potential $V(Q)$.  The slow evolution
%is needed to obtain a negative pressure,
%$p_Q=\frac{1}{2}\dot{Q}^2-V(Q)$, so that the kinetic energy density is
%less than the potential energy density.
Without modifying gravity, the most obvious way to explain this
observation is through the introduction of a quintessence field
\cite{ratrapeebles,quint,reviewde}: a time dependent scalar field
whose current homogeneous background configuration dominates the
energy density and has an equation of state which is negative.
Indeed, if the cosmic scale factor is accelerating today, the equation
of state $w \equiv p/\rho < -1/3$.

In this paper, we present a simple, yet intriguing observation: even
without specifying the details of the underlying field theory
responsible for the present quintessential behaviour of our universe,
if this theory has something to do with supersymmetry, supergravity or
(super)string theory, then one can deduce that the total energy
density must receive contributions of the order of the critical
density from the {\em interactions} between the quintessence field and
other naively ``decoupled'' fields.  Furthermore, this contribution
can generically enhance the negativity of the equation of state.  In
effect, supersymmetrizing the quintessence model can generically
enhance the negativity of the equation of state!  Of course, as we
will explain, this statement is contingent upon the assumption that
the cosmological constant is canceled by some as of yet unknown
mechanism.  Other possible effects that enhance the negativity of
equation of state in the context of supergravity can be found in
\cite{sugraquint}.

In addition, but somewhat on the flipside, we remind the reader that
supersymmetry (which arguably is the best motivated physics beyond the
Standard Model) is naturally at odds with quintessence models because
of radiative corrections induced by the nonrenormalizable terms
arising from the \Kahler potential.  Although this observation has
been made previously (see for example \cite{koldalyth,robust}), we
would like to reemphasize this important point which is relevant for
our main result of the paper.\footnote{Ref.~\cite{choi} also discusses
other aspects of general difficulties of embedding quintessence into
supergravity.}  This implies that if an equation of state for the dark
energy is observationally determined to be $0 > w > -1$ or if time
variation of the equation of state of the dark energy can be
observationally confirmed, there is a strong motivation to alter the
standard picture of supersymmetry and supergravity or to specify
special symmetries protecting the quintessence mass.  However,
whenever the traditional supersymmetric embedding of the quintessence
is possible, our main result applies.

%the description of the evolution of the
%universe is in terms of a single light field $Q$ might turn out to be
%too naive.  

The usual lores of field theory and string theory dictate that only
gauged symmetries are fundamental.  Hence, supersymmetry can only play
a fundamental role if it is a gauge symmetry.  Being a spacetime
symmetry, gauging supersymmetry produces supergravity.  Furthermore,
any fundamental supersymmetric theory such as string theory has
supergravity as its low energy effective action.  Hence any
supersymmetric embedding of quintessence is really in the context of
supergravity.  Of course, in the limit that all field amplitudes and
energies are much smaller than the Planck scale (we will denote
$M_p=2\times 10^{18}$ GeV as the reduced Planck scale), supergravity
reduces to a globally supersymmetric theory.  However, in the case of
most known quintessence models, because the field amplitudes prefer to
attain Planckian values \cite{al}, 
the supergravity structure must be taken into
account for a supersymmetrized quintessence theory.

One crucial ingredient in our observation is that in any string,
superstring, or supersymmetric theory, scalar fields which are
gravitationally coupled to all the other degrees of freedom and have
vacuum expectation values (vevs) of the order of the Planck scale
are ubiquitous.  These fields are usually required to have masses much
larger than the expansion rate $H_0$ today to have acceptable
cosmology and in practice usually have mass of order electroweak scale
of 100 GeV.  
These fields are commonly called moduli and we
collectively denote them by $\Phi$.  
%These fields commonly go under the name of moduli and we
%collectively denote them by $\Phi$.  
The important property that
the moduli have vevs of order $M_p$ reflects the fact that
supergravity has a natural scale of $M_p$.
% associated with it.

For example, in $N=1$ phenomenological supergravity models
\cite{sugra} supersymmetry (SUSY) is broken in a hidden sector and the
gravitational strength force plays the role of a messenger by
transmitting SUSY breaking to the visible sector. In these models
there exist scalar fields which are responsible for supersymmetry
breaking.  Their mass is of the order of $(10^2-10^3)$ GeV and their
coupling to the other fields is only gravitational.  Another common
example is in string derived supergravity models, all of which have
massless fields that parametrize the continuous ground state
degeneracies characteristic of supersymmetric theories.  These
fields, such as the dilaton and massless gauge singlets of string
volume compactification, are massless to all orders in perturbation
theory and can obtain their mass of order a TeV from the same
nonperturbative mechanism which breaks supersymmetry.\footnote{See 
\cite{deCarlos:1993jw}
for a discussion of other generic cosmological properties of string moduli 
fields.}

Our observation is that, even though we naively expect that only their
particle excitations can have any effect on the late time cosmological
evolution because the moduli fields $\Phi$ have masses much larger
than the Hubble rate, the very simple fact that their field amplitude
is of order $M_p$ gives any gravitational interactions of the form
\begin{equation}
\Phi^2 H_0^2,  
\label{eq:interaction}
\end{equation}
an interaction energy contribution comparable to the critical density
of the universe ($ \sim M_p^2 H_0^2$).  Furthermore, since $H_0$ is driven
by the quintessence by definition, the $\Phi$ vevs are
important for quintessence dynamics.  Since Eq.~(\ref{eq:interaction})
naturally arises from \Kahler potential couplings of $\Phi$ to the
quintessence field $Q$, such effects in supersymmetric quintessence is
generic.  Unfortunately, as we will explain, because of such
couplings, radiative corrections to the quintessence mass tend to
destabilize the requisite flatness of the quintessence
potential. (Note that any symmetries protecting the quintessence mass
still do not eliminate the interaction energy contribution.)
Since symmetries forbidding these particular radiative corrections are
rare, phenomenological viability of quintessence in the usual
supersymmetry picture is questionable.  This can be viewed as good
news in the sense that since we expect to have an observational handle on
the quintessence sector, we therefore have a new experimental probe of
the \Kahler potential.

The role of the large mass of $\Phi$ is that it causes the $\Phi$ dynamics
to decouple from the dynamics of the quintessence, leaving only the
constant vev of $\Phi$ in the interactions of the form
Eq. (\ref{eq:interaction}) relevant for the quintessence dynamics.  Of
course, in practice, the actual magnitude and the resulting equation
of state for the interaction energy is model dependent not only on the
type of coupling represented by Eq. (\ref{eq:interaction}), but the
potential of the quintessence field itself.  This will be illustrated
explicitly in this paper.

%, allowing the moduli fields to give a contribution to the
%energy density of the universe of the order of the critical energy
%density!

These conclusions in fact hold for {\it any} scalar field $\Phi$
gravitationally coupled to quintessence as long as its mass is larger
than the present-day Hubble rate and its vacuum expectation value is
of the order of $M_p$. This opens up the possibility that a
significant role in the present cosmological evolution of the universe
is played by scalar fields arising not only in supersymmetric or
(super)string theories, but also, for instance, in brane-world
scenarios.
It is encouraging that, despite the fact that one of the major
problems facing Planck scale physics is the lack of predictivity for
low-energy physics, some information on high energy physics may be
inferred indirectly through its effects on the present-day cosmological
evolution of the universe.

The rest of the paper is organized as follows.  In section 2, we give
an estimate for the energy contribution from the interaction between
the \Kahler moduli and the quintessence field.  In section 3, we argue
why the quintessence picture is typically (but not necessarily) at
odds with supersymmetry.  In section 4, we give a careful calculation
of the interaction energy starting from a generically parametrized
\Kahler potential.  Section 5 gives a different picture of the effect
of the interaction energy from an effective field theory point of
view.  It can be summarized as follows: potentials can be flattened by a field
redefinition because nonminimal kinetic term corrections are
generically expected to be large in particle physics.  We summarize
and conclude in section 6.

{\bf 2.}~~Let us first discuss the form of the moduli potential.  In
the usual nonrenormalizable hidden sector models, supersymmetry
breaking vanishes in the limit $M_p\rightarrow\infty$.  Since the
potential for a generic moduli field $\Phi$ is generated by the
same physics associated to supersymmetry breaking in the hidden
sector, its potential takes the 
%generic 
form

\begin{equation}
V(\Phi)=\widetilde{m}^2 \:M_p^2\:{\cal V}(\Phi/M_p),
\label{pot}
\end{equation}
where $\widetilde{m}\sim$ TeV is the soft supersymmetry breaking mass.
The potential for this moduli direction vanishes in the in the limit
$M_p\rightarrow\infty$ since $\widetilde{m}\rightarrow 0$ in this
limit.  The vacuum expectation value of moduli fields is naturally of
the order of the Planck scale, $\Phi_0\sim M_p$, and their excitations
around the minimum of the potential have a mass $\sim\widetilde{m}\gg
H$.  The potential (\ref{pot}) can be expanded around the minimum as
\begin{equation}
V(\Phi)=\widetilde{m}^2\left(\Phi-\Phi_0\right)^2,
\label{eq:heavymass}
\end{equation}
where we have assumed that $V(\Phi_0)$ vanishes.

We wish now to convince the reader  that the potential of the modulus field,
under the assumption that the quintessence field is dominating the energy
density of the universe,  generically 
receives contributions\footnote{The same kind of contributions
might arise during inflation (and spoil the flatness of the potential) 
\cite{reviewinf}, during preheating \cite{cpreh}, or 
 be relevant for the Affleck-Dine baryogenesis scenario 
\cite{cbario}.} 
of the form 
\begin{equation}
\Delta V(\Phi)=\frac{1}{2}\alpha\,H^2\Phi^2
\label{con}
\end{equation}
for the very same reason that
the moduli fields are coupled gravitationally to all the other
degrees of freedom and therefore to the quintessence
field as well. The coefficient $\alpha$ is of order unity and
its sign may be either positive or negative.
Let us just give an example
of one possible source of such new contributions to $V(\Phi)$. 

The Lagrangian of any scalar field in low energy supergravity is
determined by the (holomorphic) superpotential $W$ and by the
(non-holomorphic) K\"{a}hler potential $K$ \cite{sugra}.  The
K\"{a}hler potential determines the kinetic terms of the scalar fields
according to the formula
\begin{equation}
{\cal L}_{\rm kin}=\frac{\partial^2 K}
{\partial\varphi_i^*\partial\varphi_j}\partial_\mu\varphi_i^*\partial^\mu\varphi_j
\label{eq:genkin}
\end{equation}
where $\varphi_i$ are complex scalar fields (such as the modulus
$\Phi$ or the quintessence $Q$) of any SUSY multiplet.  In general the
K\"{a}hler potential is an expansion in inverse powers of $M_p$ and
contains all possible terms allowed by the symmetries of the
system.\footnote{On even more general grounds, the K\"{a}hler
potential is expected to be an expansion in inverse powers of
$\Lambda_{\rm UV}$, the ultraviolet energy cut-off of the theory.}
For instance, usually there is no symmetry forbidding the K\"{a}hler
potential to take the form
\begin{equation}
K=\Phi^*\Phi+Q^*Q+\lambda\,(\Phi^*\Phi)^m (\,Q^*Q)^n,
\label{cor}
\end{equation}
where the first two terms induce the canonically normalized kinetic
terms for the modulus and for the quintessence field, $\lambda$ is a
numerical coefficient naturally of the order of unity whose sign may
be positive or negative.\footnote{As is customary, we set $M_p=1$
whenever dimensional quantities are not being discussed.} Because the
first two terms exist for {\em any} canonically normalized \Kahler
potential, the term $\delta K=\lambda (\Phi^*\Phi)^m\,(Q^*Q)^n$ is not
forbidden by any gauge symmetries that preserve kinetic terms and
hence is expected to be there for the very simple reason that
gravitational interactions exist.

To see how Eq. (\ref{cor}) gives rise to terms of the form
Eq.~(\ref{con}), consider the equation of motion for $Q$.  The
quintessence field $Q$ rolls down a potential according to the
equation of motion $ \ddot{Q}+3H\dot{Q}+V'(Q)=0$, where $H$ is the
Hubble constant satisfying the Friedmann equation
\begin{equation}
H^2=\left(\frac{\dot{a}}{a}\right)^2=\frac{1}{3 M_p^2}\left(
\frac{1}{2}\dot{Q}^2+V(Q)+\rho_{B}\right), 
\end{equation}
where $a$ is the scale factor and $\rho_B$ is the remaining background energy
density.  Since at present the quintessence field $Q$ dominates the
energy density of the universe, we can write
\begin{equation}
\frac{1}{2}\dot{Q}^2=\frac{3}{2}\left( 1+w_Q\right)H^2M_p^2
\label{eq:qdot}
\end{equation}
and $V(Q)=\frac{3}{2}\left( 1-w_Q\right)H^2M_p^2$. Note that the mass
of the quintessence field $Q$ should naturally be of the order of the
current Hubble rate $H_0\sim 10^{-42}$ GeV.

Eqs.~(\ref{eq:genkin}), (\ref{cor}), and (\ref{eq:qdot}) give rise
to a new contribution 
to the modulus potential of the form (\ref{con}) with
\begin{equation}
\alpha \sim 3 \lambda (1+w_Q) \left(\frac{\Phi^* \Phi}{M_p^2}\right)^{(m-1)} 
\left(\frac{Q^* Q}{M_p^2}\right)^{(n-1)}
\end{equation} 
where we have purposely been careless about factors of 2 in the
kinetic normalization\footnote{We have not been careful in treating
$Q$ and $\Phi$ consistently as real or complex fields when they should
be consistently complex.}  (we do a careful analysis in section 4).
Since both $Q$ and $\Phi$ typically have vevs near $M_p$, the
coefficient $\alpha$ is generically not suppressed.

We stress again that this is only one of the possible new
contributions to the potential of the modulus. All of them are
expected to be parametrized by the expression (\ref{con}).  For
example, this type of contribution may arise from nonminimal coupling
to gravity of the form
\begin{equation}
\xi R \Phi^2
\end{equation}
where $R$ is the Ricci scalar and $\xi$ is a constant.  Note that this
term is generated at one loop order even when absent at tree level.
(Terms of the form in Eq.~(\ref{con}) arising from nonminimal coupling
without any reference to supersymmetry have been utilized, for
example, by \cite{faraoni}.)
%Minimizing the modulus potential, we find that the vacuum expectation value
%of the modulus field has been dynamically shifted from its value $\Phi_0$
%\begin{equation}
%\langle\Phi\rangle=\frac{\widetilde{m}^2}{\widetilde{m}^2+\alpha H^2}\,\Phi_0.
%\label{pp}
%\end{equation}
%Analogously, the modulus potential at the new minimum is given by
%\begin{equation}
%\langle V\rangle=\frac{\alpha}{2}\frac{\widetilde{m}^2\,H^2}{\widetilde{m}^2+
%H^2}\,\Phi_0^2\simeq \frac{\alpha}{2}\,H^2\,\Phi_0^2.
%\end{equation}
Since 
%the vev 
$\langle \Phi \rangle =\Phi_0$ is naturally of the order
of $M_p$, we find that the interaction energy contributes
\begin{equation}
\langle V + \Delta V \rangle \sim H^2 M_p^2
\end{equation}
while because of the large mass $\widetilde{m}$, the small shift in
the $\Phi$ vev is negligible. 
% We conclude that the unavoidable
%gravitational interaction of the moduli fields with the quintessence
%field shifts by a tiny amount the vacuum expectation value of the
%modulus field. Since the potential is very curved around the minimum,
%$V''(\langle\Phi\rangle)\sim \widetilde{m}^2\gg H^2$, the modulus
%field is strongly captured in the minimum of its potential. When the
%quintessence field rolls down its potential, the modulus field moves
%adiabatically to its new time-dependent equilibrium value and
%$\langle\dot\Phi\rangle\ll \dot Q\sim H M_p$.  This adiabatic change
%results in a contribution to the energy density of the universe
%comparable to the critical energy density of the universe today!

{\bf 3.}~~Note that the example of Eq.~(\ref{cor}) with $m=n=1$ is
already phenomenologically unsatisfactory because there is an one loop
mass contribution from the resulting effective Lagrangian term
\begin{equation}
\lambda |Q|^2 |\partial \Phi|^2 
\end{equation}
that is too large for the quintessential behavior to be maintained.
Even in the most optimistic scenario, we expect this coupling (with
the attendant partial cancellation contributions from SUSY partners)
to generate quintessence mass corrections of at least of order
\begin{equation}
\delta m_Q \sim \frac{\sqrt{\lambda}}{4\pi} \left(
\frac{\widetilde{m}}{M_p} \right) \widetilde{m}
\end{equation}
which for $\widetilde{m}\sim $TeV yields $\delta m_Q \sim 10^{-5}$ eV.
Although this is not necessarily disastrous for the quintessence
energy, it then becomes difficult to explain why the quintessence has
not settled to its minimum already thereby making the quintessence
energy contribution more like a cosmological constant.

Indeed, that is why typically the quintessence mass is required to
be of the order
\begin{equation}
m_Q \sim H_0\sim 10^{-33} \mbox{eV}
\end{equation}
which is indeed a very tiny mass scale compared to any other mass
scales that have been measured experimentally.  The fact that any tiny
effect can destabilize this tiny mass scale makes quintessence a very
sensitive probe of the \Kahler potential.  It is important to keep in
mind that because the first two terms in Eq.~(\ref{cor}) are gauge
invariant and always present, one cannot eliminate the term
proportional to $\lambda$ simply by using gauge symmetries that act
only on the kinetic term.  Furthermore, experience with the Standard
Model has taught us that any terms not forbidden by fundamental
symmetries always exist in the Lagrangian.  Unless symmetry principles
can be found to eliminate the generic nonminimal term in
Eq.~(\ref{cor}) or symmetry principle cancels the radiative
corrections coming from these nonminimal terms exactly even in the
presence of SUSY breaking, observational confirmation of the
quintessence picture may make the standard picture of supersymmetry
quite unfavorable.\footnote{One possibility for protecting the
quintessence mass is through a global symmetry which can realize
quintessence as a pseudo-Nambu-Goldstone boson.  Model building along
these lines has been considered, for example, in \cite{choi,kim}.}

For the rest of the paper, we will assume that the required symmetry
exists to protect the quintessence mass.  Hence, in the next section we
will solve the general problem with the \Kahler potential of
%the form in
 Eq.~(\ref{cor}) where $m$ and $n$ are natural numbers not
necessarily equal to 1.

{\bf 4.}~~We devote this section to a more complete analysis of the
dynamics of the system made of the quintessence and the moduli
fields. As we shall see, this detailed analysis confirms the
conclusions in the previous sections.  The main objective of the
analysis is to compute the equation of state with the interaction
energies taken into account.  The choice of the toy model will be
based on the aim of demonstrating the natural existence of the
enhancement of the negativity of the equation of state rather than
complete generality.  As we advertised previously, although the exact
numerical value of the energy and pressure contribution due to the
interaction is sensitive to the details of the quintessence potential,
its order of magnitude is not.

Let us consider the generic action
\begin{equation}
S_{\rm M}=\int d^{4}x\,\sqrt{-g}\left[G^{i}_{\, j}D_{\mu }\phi _{i}D^{\mu }
\phi ^{j*}+e^{G}(3-G_{i}(G^{-1})^{i}_{\, \, j}G^{j})\right], 
\end{equation}
where $G=K+{\rm ln}\,|W|^2$, $ G^{i} \equiv \partial _{\phi
_{i}}G$, $G_{i} \equiv \partial _{\phi ^{i*}}G$, and $G^{i}_{\, \, j}
\equiv \partial _{\phi _{i}}\partial _{\phi ^{j*}}G$.  We have set the
reduced Planck constant $M_p=1$.  Choosing the a K\"ahler potential of
the form Eq. (\ref{cor}) we find the kinetic terms to be
\begin{eqnarray}
\label{eq:toyaction}
S_{\rm kin} & = & \int d^{4}x\sqrt{g}g^{ab}\left[(1+\lambda n^2 |\Phi |^{2
m} |Q|^{2n-2} )
\partial _{a}Q \partial _{b}Q ^{*}+(1+\lambda n^2 |Q|^{2n} |\Phi|^{2m-2})
\partial _{a}\Phi \partial _{b}\Phi ^{*} \right. \nonumber \\ 
& & \left. +\lambda n m |\Phi|^{2m-2} |Q|^{2n-2} \left( \Phi ^{*}Q 
\partial _{a}\Phi \partial _{b}Q ^{*}+ \Phi 
Q^{*}\partial _{a}\Phi ^{*}\partial _{b}Q \right)\right].
\label{action}
\end{eqnarray}
As for the potential, we will assume that there is a contribution to
the effective potential of the form\begin{equation}
V(\Phi)=\widetilde{m}^{2}\left|\Phi -\Phi _{0}\right|^{2},
\end{equation}
where \( \Phi _{0} \) is of order \( M_{p} \). For the quintessence,
we also add by hand a potential $V(Q)$ that leads to a negative
equation of state and energy density order of the critical density
today.  Based on the criterion of ease of mathematical manipulation,
we choose $V(Q)=V_0 \, e^{-\beta R}$ where
$R=\frac{1}{2}\left(Q+Q^*\right)$ \cite{quinteexp}.\footnote{This type
of potential is phenomenologically undesirable for couple of reasons.
One is that it does not give any potential to the imaginary part of
$Q$.  Another is that big bang nucleosynthesis bounds make this
potential undesirable \cite{quinteexp}.  Nonetheless, since we
are not concerned with the global behavior but rather the local
behavior, these choices should suffice to illustrate the effect of
interest.}  We neglect the background energy density contributions
such as those from cold dark matter, baryons, and radiation since we
are not interested in the global tracking properties but more on local
properties.  The qualitative aspects of the present demonstration
should not depend upon the details of these choices as supported by
the general arguments given by the previous sections.  Furthermore, we
justify not specifying the details of the superpotential by the fact
that we do not know how most of the cosmological constant is
cancelled.  The kinetic term effect that we analyze here is likely to
be unaffected by the solution to the cosmological constant problem as
long as the solution to the cosmological constant problem does not
involve derivatively coupled terms.  Finally, we will assume that the
quintessence energy density dominates and will neglect the background
matter and radiation energy density.  This is justified since we are
concerned with local properties without worrying about tracking
behavior.

The equations of motion for the modulus and for the quintessence field
reads\begin{eqnarray}
\frac{1}{a^{3}}\partial _{t}(a^{3}\partial _{t}\Phi )+ \lambda \frac{m}{a^{3}}(Q^{*})^{n}(\Phi ^{*})^{m-1}\partial _{t}(a^{3}\partial _{t}(Q^{n}\Phi ^{m}))+\widetilde{m}^{2}(\Phi -\Phi _{0}) & = & 0\\
\frac{1}{a^{3}}\partial _{t}(a^{3}\partial _{t}Q)+ \lambda \frac{n}{a^{3}}((Q^{*})^{n-1}(\Phi ^{*})^{m}\partial _{t}(a^{3}\partial _{t}(Q^{n}\Phi ^{m}))-\frac{\beta }{2}V_{0}e^{-\beta R} & = & 0
\end{eqnarray}
Note that there is a separation of scales in that \( \widetilde{m}\gg
H \).  Hence, we will define the perturbation order bookkeeping
variable \( s \) that reflects this hierarchy. In other words, we
introduce a homogeneous perturbation $\phi(t)$ about the constant vev
(which solves the equation of motion in the limit
$\widetilde{m}\rightarrow \infty$) as \begin{equation} \Phi =\Phi _{0}+s \phi(t) \end{equation}
and expand everything to first order in \( s \).

Furthermore, since the \Kahler expansion is uncontrolled for \begin{equation}
\lambda (\Phi ^{*}\Phi )^{m}(Q^{*}Q)^{n}>1\end{equation}
 we will impose a hierarchy for the computational sake that\begin{equation}
\label{eq:vevhierarchy}
\lambda \Phi _{0}^{2(m-1)}Q_{0}^{2n}\sim \lambda \Phi _{0}^{2m}Q_{0}^{2(n-1)}\sim \lambda \Phi _{0}^{2m-1}Q^{2n-1}\sim O(1/10)
\end{equation}
 where the \( Q_{0} \) is the zeroth order solution with \( \lambda =0 \).
To be conservative, we will treat the perturbation in \( \lambda  \)
on the same order as perturbation in \( s \) and assign a bookkeeping
device \( r \) for the order of \( \lambda  \):\begin{equation}
Q=Q_{0}(t) +rq(t)\end{equation}
where \( q \) is the homogeneous perturbation. As we will see, even
then, the \( \Phi  \) perturbation to first order in \( s \) becomes
unimportant for the energy and the equation of state. Finally, we
also expand the expansion rate \( H=\dot{a}/a \) as a perturbation
series in \( r \) as\begin{equation}
H=H_{1}(t)+rh(t)\end{equation}
where \( h \) is the homogeneous perturbation.

The expansion to zeroth order in \( r \) and \( s \) gives rise
to the usual quintessence equations of motion\begin{equation}
\ddot{Q}_{0}+3H_{1}\dot{Q}_{0}-\frac{\beta }{2}V_{0}e^{-\beta R}=0\end{equation}
\begin{equation}
H_{1}^{2}=\frac{1}{3}\left( |\dot{Q}|^{2}+V_{0}e^{-\beta R}\right) \end{equation}
which have the cosmological solutions\begin{equation}
Q_{0}=\frac{2}{\beta }\ln \left(\frac{t}{\tau }\right)\end{equation}
\begin{equation}
H_{1}=\frac{4}{\beta ^{2}t}\end{equation}
\begin{equation}
\label{eq:V0andtau}
V_{0}=\frac{4(12-\beta ^{2})}{\beta ^{4}\tau ^{2}}
\end{equation}
where we must keep in mind that \( Q_{0} \) has to be chosen to satisfy
Eq. (\ref{eq:vevhierarchy}). Since \( \beta \sim O(1) \), we then
should choose \( t/\tau \sim O(1) \).

The equations of motion to first order in \( r \) and \( s \) yield
\begin{eqnarray}
\ddot{\phi }+3H_{1}\dot{\phi }+\lambda mnQ_{0}^{2n-1}\Phi
_{0}^{2m-1}(\ddot{Q}_{0}+3H_{1}\dot{Q}_{0})+ & & \nonumber \\
\lambda mn(n-1)Q_{0}^{2n-2}\Phi
_{0}^{2m-1}\dot{Q}_{0}^{2}+\widetilde{m}^{2}\phi  & = & 0 \\
\ddot{q}+3h\dot{Q}_{0}+3H_{1}\dot{q}+\lambda
n^{2}(n-1)Q_{0}^{2n-3}\Phi _{0}^{2m}\dot{Q}_{0}^{2}+ & & \nonumber \\
\lambda n^{2}Q_{0}^{2n-2}\Phi
_{0}^{2m}(\ddot{Q}_{0}+3H_{1}\dot{Q}_{0})+\frac{\beta
^{2}}{4}(q+q^{*})V_{0}e^{-\beta R} & = & 0 \\
\frac{1}{3}\left[ \lambda n^{2}Q_{0}^{2n-2}\Phi _{0}^{2m}\dot{Q}_{0}^{2}+\dot{Q}_{0}(\dot{q}+\dot{q}^{*})-\frac{\beta }{2}V_{0}e^{-\beta Q_{0}}(q+q^{*})\right]  & = & 2hH_{1}
\end{eqnarray}
where we have assumed \( |\frac{\beta }{2}(q+q^{*})|<1 \). As for
the boundary conditions to these perturbation equations, we
set
\begin{eqnarray} 
\phi (t_{i})=0 &  & \dot{\phi }(t_{i})=0\\
q(t_{i})=0 &  & \dot{q}(t_{i})=0
\end{eqnarray}
although any parts of the perturbation solutions that depend on the
boundary condition tend to die away faster than the non-boundary
condition dependent terms (sourced part of the solution), and therefore
is not important unless \( V_{0}\rightarrow 0 \).

Using the usual one dimensional Green's function technique, one can
solve these equations. We find for the perturbation to \( \Phi _{0} \)
the solution\begin{equation}
\phi (t)=\frac{-\lambda mn\Phi _{0}^{2m-1}}{\tilde{m}t^{2}}
\left(\frac{2}{\beta }\right)^{2n}\left[ \tilde{x}
\left(\ln \left[\frac{t}{\tau }\right]\right)^{2n-1}+(n-1)\left(
\ln \left[\frac{t}{\tau }\right]\right)^{2n-2}\right] +\textrm{b}.\textrm{c}.\textrm{ dep}.\textrm{ terms}\end{equation}
 where \( \tilde{x}\equiv 12/\beta ^{2}-1 \) and the {}``b.c. dep.
terms'' represent boundary condition dependent subleading terms that
die away as a function of time if \( \tilde{x}>0 \).
For the perturbation to \( Q_0 \), we have\begin{eqnarray}
q(t) & = & \frac{-\lambda n^{2}\Phi _{0}^{2m}}{\beta }
\left( \frac{2}{\beta }\right) ^{2n-2}\left[ \frac{2\tilde{x}+1}{\tilde{x}}
\left(\ln \left[\frac{t}{\tau }\right]\right)^{2n-2}+\right. \nonumber
\\
 &  & \left. (2n-2)!\tilde{x}\sum _{l=0}^{2n-3}(\tilde{x}^{l-2n}+
\sum _{y=1}^{2n-l-1}2\tilde{x}^{-y})\frac{(-1)^{l}
(\ln [\frac{t}{\tau }])^{l}}{l!}\right] +
\textrm{b}.\textrm{c}.\textrm{ dep}.\textrm{ terms}
\end{eqnarray}
 where the {}``\( \textrm{b}.\textrm{c}. \) dep. terms'' again
indicate boundary condition dependent terms which die away if \(
\tilde{x}>0 \).\footnote{Note
that the singularity at \( \beta ^{2}=12 \) is an artifact of neglecting
the boundary condition dependent terms and not a true singularity.
In the limit that \( \beta ^{2}\rightarrow 12 \), Eq. (\ref{eq:V0andtau})
forces \( V_{0}\rightarrow 0 \), in which case the only source term
is the background solution sensitive to the boundary conditions. In
that case, boundary condition dependent terms naturally become important.}

\begin{figure}
{\centering \epsfbox{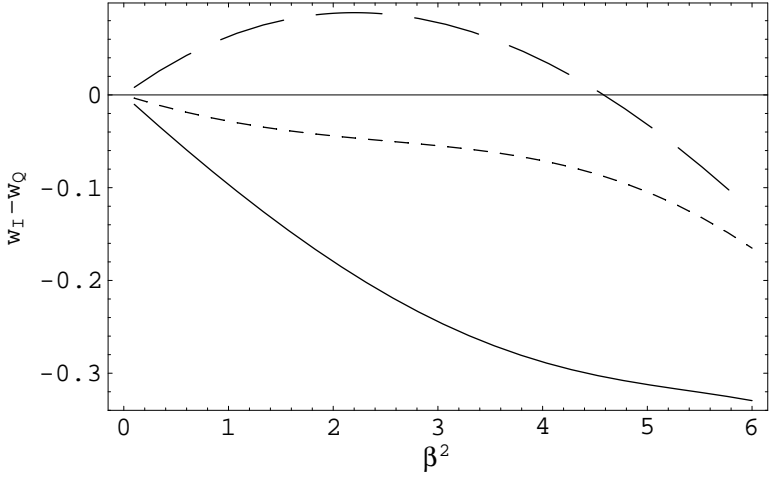} \par}

\caption{Plot of \protect\( w_{I}-w_{Q}\protect \) 
as a function of \protect\( \beta ^{2}\protect \)
for \protect\( n=2\protect \). The long dashed curve corresponds
to \protect\( t/\tau =1\protect \), the short dashed curve corresponds
to \protect\( t/\tau =5\protect \), and the solid curve corresponds
to \protect\( t/\tau =10\protect \).}
\end{figure}

Note that the validity of this solution is only in the regime when
\( |\beta q|<1 \). The resulting energy density and pressure can
be calculated to be\begin{equation}
\rho =\rho _{Q}+\rho _{I}\end{equation}
\begin{equation}
p=p_{Q}+p_{I}\end{equation}
\begin{equation}
\rho _{Q}\equiv \frac{48}{\beta ^{4}t^{2}}\end{equation}
\begin{equation}
p_{Q}\equiv \frac{8(\beta ^{2}-6)}{\beta ^{4}t^{2}}\end{equation}
\begin{equation}
\rho _{I}\equiv 2\lambda n^{2}
\frac{(1+\tilde{x})^{2n-1}}{\tilde{x}^{2n-2}}\left(\frac{2}{\beta }\right)^{2n}
\frac{\Phi _{0}^{2m}}{t^{2}}\left( \frac{\tau }{t}\right)^{\tilde{x}/
(1+\tilde{x})}\Gamma \left(2n-1,\frac{-\tilde{x}}{1+\tilde{x}}\ln 
[\frac{t}{\tau }]\right)\end{equation}
\begin{equation}
p_{I}\equiv -2\lambda n^{2}\frac{\tilde{x}\Phi _{0}^{2m}}{t^{2}}(\frac{2}{\beta })^{2n}\frac{\tau }{t}\Gamma (2n-1,-\ln \frac{t}{\tau })\end{equation}
 where \( \rho _{I} \) and \( p_{I} \) represent interaction energy
and pressure, and \( \Gamma (a,x) \) is the incomplete gamma function.\footnote{The
incomplete gamma function is defined as \( \Gamma (a,x)\equiv \int _{x}^{\infty }dtt^{a-1}e^{-t} \).} 

First, note that since we are working in the regime \( \tilde{x}>0 \)
and \( \Gamma >0 \), the signs of \( \rho _{I} \) and \( p_{I} \)
take the signs of \( \lambda  \) and \( -\lambda  \), respectively.
This means that for \( \lambda >0 \), the pressure contribution is
negative while the energy contribution is positive. Secondly, we can
write the equation of state to leading order in \( \lambda  \) as\begin{equation}
w=\frac{p}{\rho }=w_{Q}+(w_{I}-w_{Q})\Delta \end{equation}
\begin{equation}
w_{Q}\equiv \frac{p_{Q}}{\rho _{Q}}=-1+\frac{\beta ^{2}}{6}<0\end{equation}
\begin{equation}
w_{I}\equiv \frac{p_{I}}{\rho_{I}}=-\left( \frac{\tilde{x}}{1+\tilde{x}}
\right) ^{2n-1}\left(\frac{\tau }{t}\right)^{\frac{1}{1+\tilde{x}}}\frac{\Gamma (2n-1,-\ln \frac{t}{\tau })}{\Gamma (2n-1,\frac{-\tilde{x}}{1+\tilde{x}}\ln \frac{t}{\tau })}<0\end{equation}
\begin{equation}
\Delta \equiv \frac{\rho _{I}}{\rho _{Q}}\end{equation} where \(
\textrm{sign}(\Delta )=\textrm{sign}(\lambda ) \) and \( w_{I} \) is
the interaction equation of state. Hence, if \( \lambda >0 \) and \(
w_{I}-w_{Q}<0 \) (equivalently \( |w_{I}|>|w_{Q}| \)), the total
equation of state becomes more negative. Since \( t/\tau \sim O(1) \),
we plot in Fig. 1, \( (w_{I}-w_{Q}) \) as a function of \( \beta^2 \)
for \( t/\tau =1, 5,
% \mbox{and} 
10 \) with \( n=2 \). It clearly shows that a negative contribution to
the equation of state is quite generic.  Finally, note that the power $m$
of the modulus field $\Phi$ enters the total equation of state $w$ only
through the relative energy ratio $\Delta$.

{\bf 5.}~~Although we have focused on the particular form of the
\Kahler potential and a perturbative analysis above, the fact that the results are more
general can be seen simply as follows.  First, note that since the generic
``heavy'' fields which we have denoted as $\Phi$ are much more massive than
the quintessence field $Q$, the field $\Phi$ can be integrated out
leaving only the participation of a {\em constant} $\Phi$ vev in the
dynamics of the quintessence: i.e. we can set $\dot{\Phi}=0$.  In that
case, the action for the quintessence (say for the real part of $Q$) looks like
\begin{equation}
S_{eff}= \int d^4 x \sqrt{g} \left[ (1+f(Q)) \frac{\dot{Q}^2}{2} - V(Q) \right]
\end{equation}
where $f(Q)$ is a function that has only a parametric dependence on
$\langle \Phi \rangle$ which in the SUGRA case comes from the
nonminimal \Kahler potential.  Now, a field redefinition
\begin{equation}
\tilde{Q}=\int dQ \sqrt{1+f(Q)}
\end{equation}
puts $S_{eff}$ into the form
\begin{equation}
S=\int d^4 x \sqrt{g} \left[\frac{\dot{\tilde{Q}}^2}{2} - V(Q(\tilde{Q}))  \right]
\end{equation}
where now $V(Q(\tilde{Q}))$ as a function of $\tilde{Q}$ can be a much flatter potential than what
one originally considered to be the potential $V(Q)$.  The
flatness of the potential is characterized by the smallness of
\begin{equation}
\delta  = M_p \frac{d V(Q(\tilde{Q}))}{d \tilde{Q}} = M_p \frac{d
V(Q)}{d Q} \frac{1}{\sqrt{1+f(Q)}}
\end{equation}
which explicitly shows that if $f(Q)$ is large and positive, there is
a flattening of the potential which, in turn, translates to an
enhanced negativity of the equation of state.  Hence, the
``interaction energy'' discussed in the previous section can be seen
as the flattening of the potential in a different field variable.

Explicitly, suppose the ``second slow roll condition''
\begin{equation}
M_p^2 \frac{d^2 V}{d \tilde{Q}^2}/V < 1 
\end{equation}
is satisfied.  Then, we can write the equation of state for the
quintessence as
\begin{equation}
w = 1- \frac{2}{1+ \frac{1}{6} \delta^2 }
\end{equation}
which shows that as $f(Q)\rightarrow \infty$, $\delta \rightarrow 0$,
$w \rightarrow -1$.  Hence, we expect the results of the previous
section regarding the enhancement of the equation of state to hold for
more general type of \Kahler potentials.  Furthermore, we expect the
upper bound on the enhancement to be $w =-1$.

Note also that although the \Kahler potential of the form
\begin{equation}
K= -\ln(Q+\bar{Q}) 
\end{equation}
frequently encountered in string theory does not have the form of a
polynomial with a single term dominating for all field values,
locally, away from the singular points of the $\ln$ function, the
potential can always be expanded in Taylor series with one term as the
leading dominant term.  Explicitly, we can rewrite this \Kahler
potential up to \Kahler transformation dependent terms as
\begin{equation}
K= k \bar{k} -  k \bar{k} ( k+ \bar{k})+ ...
\end{equation}
where
\begin{eqnarray}
k & \equiv & Q(t_0) + (Q(t_0)+ \bar{Q}(t_0)) k \\
\bar{k} & \equiv & \bar{Q}(t_0) + (Q(t_0)+ \bar{Q}(t_0)) \bar{k}
\end{eqnarray}
with $Q(t_0)$ being the value of the quintessence field at some point
in time (say today) and $k$ being the dynamical variable.  If one
restricts to real values of $k$ as we have done in the previous
section, similar results will follow.

Finally, we would like to emphasize one of the main weaknesses of the present
paper which lies in assuming the existence of matching of the
quintessence models to the potential involving a nonminimal \Kahler
potential.  More concretely, what we are stating is that if we start
with 
\begin{equation}
S= \int d^4 x \sqrt{g} \left[ (1+f_0 (Q, \Phi)) \frac{\dot{Q}^2}{2} -
V_0(Q,\Phi) \right]
\end{equation}
where $\Phi$ is dynamical (before integrating them out) and try to
match to an ansatz quintessence potential $V_A(Q)$ after integrating
out $\Phi$, we are matching $V_0(Q,\langle \Phi \rangle)= V_A(Q)$.
Although in some sense, this matching is arbitrary, it is not
unnatural.  Furthermore, as we stated before, because we do not know
the cancellation mechanism of the cosmological constant, it is
difficult to address this assumption more rigorously.  Even if we
relaxed the matching assumption such that we would not know the final
fate of the effect of the generically nonminimal \Kahler potential,
the fact that the nonminimal kinetic term which induces
$\dot{Q}$-$\Phi$ interaction energies of the order $H^2 M_p^2$ can
flatten the potential leading to a negative contribution to the
equation of state is still a true and interesting statement.

{\bf 6.}~~In this brief paper we have pointed out that any scalar
field with gravitational coupling to quintessence and vacuum
expectation values of the order of the Planck scale play a significant
role in the present-day cosmological evolution of the universe. Our
findings suggest that in a realistic particle theory approach to the
dark energy problem, the use of a single quintessence field for models
is likely to miss significant contributions to the negative equation
of state.  Another way of viewing this is that the nonminimal kinetic
terms generically expected in realistic particle theories can imply a
significantly flatter (or steeper) potential than what one would write
down without knowing that nonminimal kinetic terms were generic.  If
the potential is flatter, then the equation state is more negative,
allowing certain naively ruled out models to be revived.  Furthermore,
we have argued that supersymmetric embedding of the quintessence is
generically difficult because of sensitivity of the quintessence mass
to generic terms in the \Kahler potential.  This presents the exciting
possibility that confirmation of quintessential picture may lead to
new probes into the underlying high energy physics.
%In summary, 
Our
observations open up new possibilities, such as testing high energy
physics through its effects on the cosmological evolution of the
universe or significantly changing the quintessential phenomenology,
allowing naively ruled out quintessence parameter space to become viable.
\vskip1cm

\section*{Acknowledgments}
We would like to thank J. Cline, T. Dent, G. Gabadadze, and
R. Rattazzi for helpful discussions.

\end{document}